          [2001/04/25 1.1 (PWD)]
\documentclass[a4paper,twocolumn]{esapub} 
\usepackage{natbib}
\usepackage{graphics,graphicx}
\usepackage{here}

\title{A search for cyclotron resonance features with {\it INTEGRAL}}
\author{Y. Okada}
\author{H. Niko}
\author{M. Kokubun}
\author{K. Makishima}
\affil{Department of Physics, School of Science, University of Tokyo, Tokyo, Japan, yokada@amalthea.phys.s.u-tokyo.ac.jp}
\author{M. Nakajima}
\author{T. Mihara}
\author{Y.Terada}
\affil{RIKEN, Wako, Saitama, Japan}
\author{F. Nagase}
\affil{ISAS/JAXA, Sagamihara, Kanagawa, Japan}
\author{Y. Tanaka}
\affil{MPE, Garching, Germany}

\begin{document}


\maketitle

\begin{abstract}
We present an {\it INTEGRAL} observation of the Cen-Crux region
in order to search the electron cyclotron resonance scattering features
from the X-ray binary pulsars.
During the AO1 200~ks observation, we clearly detected
4 bright X-ray binaries, 1 Seyfert Galaxy, and 4 new sources
in the field of view.
Especially from GX301-2, the cyclotron resonance feature
is detected at about $\sim$ 37~keV, and width of 3--4~keV.
In addition, the depth of the resonance feature strongly
depends on the X-ray luminosity.
This is the first detection of luminosity dependence
of the resonance depth.
The well-known twin pulsars are spatially separated
by JEM-X and IBIS/ISGRI, and pulse periods are
measured individually; 296.90~sec for 1E1145-6141
and 292.5~sec for 4U1145-619.
The cyclotron resonance feature is marginally detected from 1E1145.1-6141.
Cen X-3 was very dim during the observation and
poor statistics disable us to detect the resonance features.
\end{abstract}

\section{Introduction}
Neutron stars~(NS) are regarded as having extremely strong
magnetic fields~(MFs) reaching 10$^{12-13}$ G. Nevertheless, the
origin and evolution of such strong MFs have been standing as a
long mystery. The electron cyclotron resonance scattering features
(CRSFs) in the X-ray spectrum are the only direct method to
measure the MFs on the NS. The center {\it E}$_{\rm c}$
 of CRSF in their spectra
reflects the magnetic field {\it B}, given as
$E_{\rm c} = 11.6 \times (B/10^{12}G \times (1+z)^{-1} {\rm keV}$
. In early 1990's, {\it Ginga} (Japanese 3rd X-ray satellite)
provided a powerful mean to search for CRSFs in the 2--30~keV range
in the X-ray binary pulsars~(XBPs) spectra. In addition to two already
known XBPs, new CRSFs were discovered from more than 6~XBPs.
The MF strength of XBPs seems to be
concentrated in a narrow range of B=~(1--5) $\times$ 10$^{12}$~G~(Makishima et al. 1999).
However, the high-field side of the distribution, although augmented by
{\it SAX} and {\it RXTE}, is still subject to considerable selection effects.
{\it INTEGRAL}, with its wide dynamic range and good sensitivity, is expected to be  very
powerful to search  for the CRSF  at higher energy range.
Here we report the {\it INTEGRAL} observation of 4 Galactic XBPs,
in an attempt to detect CRSFs at higher energy range of 10--150~keV.

\begin{figure}
\centering
\includegraphics[width=1.0\linewidth]{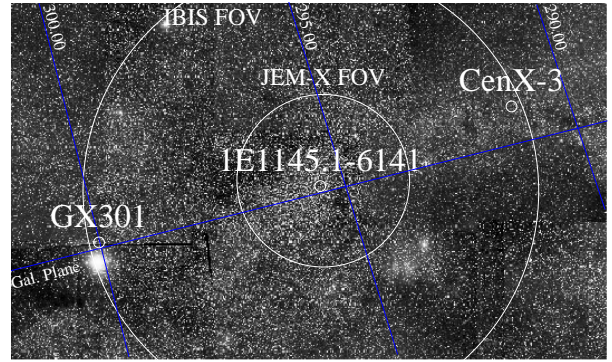}
\caption{An IBIS and JEM-X fully coded field of view superposed on the DSS image in the Cen-Crux region.}
\label{fig:single}
\end{figure}

\section{Observation and data reduction}
We have observed the Cen-Crux region during the AO1 phase,
 from the
end of June 2003 to the beginning of July 2003, with an exposure time of
$\sim$ 200~ksec, which is composed of 52 science window.  Cen-Crux
region is located on the Galactic plane and 60$^{\circ} $ apart from
the Galactic center, and it is known as a region where the bright
X-ray pulsars are concentrated.  The observations field of view
are centered near 1E1145.1-6141, one of the bright X-ray pulsar, shown 
in figure 1.  Imaging
analysis is performed using the current version~(3.0) of the Offline
Science Analysis~(OSA) software.  Spectral extraction is performed
independently for every science window.  
The data of IBIS/ISGRI are
mainly studied and we also use JEM-X and SPI spectra.

\section{Hard X-ray source population in Cen-Crux region}
We
generated the significance image of 15--40~keV and 40--100~keV range
and search for sources which exceeds 7$\sigma$ significance in
15--40~keV, this detection limit corresponds to $\sim$ 5mCrab.  As
a result, we detected 9 hard X-ray sources in the FOV~(fig. 2). 
As expected,
4 sources are well known XBP; Cen X-3, GX301-2, 1E1145.1-6141, and
4U1145-619.  NGC4945, which is one of the brightest Seyfert 2 galaxy
in hard X-ray band, is detected in the north-east region.  4 new
sources are also discovered in the FOV.  The source list is summarized
in table 1.
\begin{figure}
\centering
\includegraphics[width=1\linewidth]{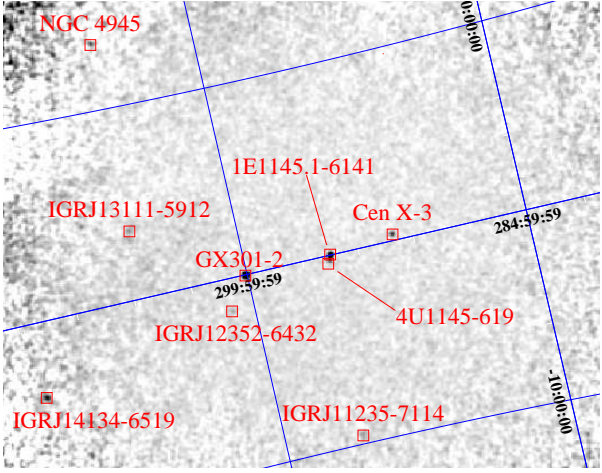}
\caption{The mosaic image of Cen-Crux region obtained by IBIS/ISGRI
in the energy range 15--40~keV. The total exposure time is $\sim$ 200~ksec.}
\label{fig:single}
\end{figure}

\begin{table*}
  \begin{center}
    \caption{List of sources detected during observations of Cen-Crux region.}
    \vspace{1em}
    \begin{tabular}[h]{cccccc}
      \hline
      Object & RA & Dec & Significance(soft/hard)$^{\ast}$ & Flux [mCrab]$^{\ast\ast}$ & Class \\
      \hline
	GX301--2 & 186.6587 & -62.7752 & 233.3/16.1 & 78 & XBP\\
	Cen X-3  & 170.3223 & -60.6268 &  26.2/$<$3 & 15 & XBP\\
	1E1145.1-6141 & 176.8983 & --61.9977&  74.9/26.1 & 50 & XBP\\
	4U1145-619& 176.9178 & -62.3190 &  28.5/10.0 & 37 & XBP\\ 
	NGC4945  & 196.3826 & -49.5000 &  6.13/ 7.7 & 30 & Seyfert 2\\ 
	IGRJ14134-6519 & 213.4368 & -65.3282 &  10.2/3.2 & 8  & ?\\
	IGRJ12352-6432 & 188.8395 & -64.5357 &  8.5/$<$3 & 12 & ?\\
	IGRJ13111-5912 & 197.8113 & -59.2058 &  7.9/4.5  & 8  & ?\\
	IGRJ11235-7114 & 170.9609 & -71.2420 &  7.0/$<$3 & 8  & ?\\
      \hline \\
      \end{tabular}
    \label{tab:table}
  \end{center}
\vspace{-5mm}
$^{\ast}$~:~ 15--40~keV for soft , 40--100~keV for hard band respectively.\\
$^{\ast\ast}$~:~15--100~keV 
\end{table*}

\section{GX301-2}
The high mass X-ray binary pulsar GX301-2 is a wind-fed neutron star
around the supergiant companion Wray~977~(Sato et al. 1986).
The pulse period is 675.7 sec~(Pravdo et al. 1995) and orbital phase is 41.508~$\pm$ 0.007 days (Sato et al. 1986).
The CRSF is firstly reported at  35.6$\pm$ 1.6~keV by {\it Ginga} (Mihara 1995, Makishima et al. 1999), 
although the detection is not firmed.
A 42.4$\pm 3.8$~keV CRSF is reported by {\it RXTE} (Coburn  et al. 2002). 
 GX301-2 was very bright ($\sim$ 100mCrab) during the observations, and
was clearly detected by JEM-X, IBIS/ISGRI, and SPI.

\subsection{Crab-ratio spectrum}
Because the CRSF detection strongly depends on the continuum spectrum,
the accuracy of the detector response matrix is very important in the
study.  However, as reported in Goldwurm et al. 2003, the response
matrix of the ISGRI detector is under development.
To make the  spectrum fitting using the response matrix distributed 
by Integral Science Data Center~(ISDC), 
a large systematic error, about 10\%, is required. 
That makes our CRSF search difficult.
In order to avoid large uncertainty caused by the current response matrix,
we firstly try the Crab-ratio spectrum analysis.
By dividing our data by the well-known power law spectrum, 
it becomes free from the complexity of the detector 
response matrix, and we can directly investigate the CRSFs.

\begin{figure}
\centering
\includegraphics[width=1\linewidth,angle=270]{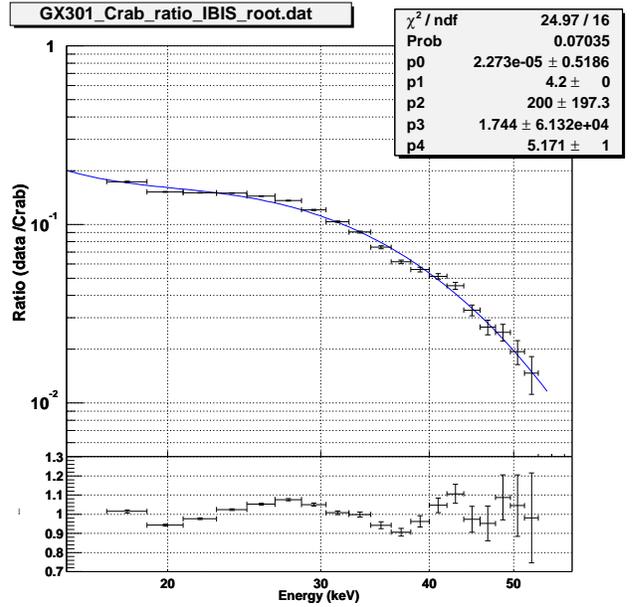}
\caption{The ISGRI Crab ratio fitting of GX301-2.}
\label{fig:crab_ratio}
\end{figure}

Figure \ref{fig:crab_ratio}~ shows the crab-ratio spectrum of ISGRI.
We fit the data for the simple continuum and obtained the chi-squared.
As shown in the bottom panel, the residual is clearly seen at about
30-40~keV.  Therefore we add the CRSF absorption model as an
additional component.  As a result, CRSF at 36$\pm$ 1.5~keV
is detected, and the detection significance of the 
CRSF is about 2.1$\sigma$. 
The absolute intensity of the CRSF is not determined,
because we used the spectrum ratio.

\subsection{Spectrum fitting}
Although the response matrix of ISGRI is still under development,
we can fit the data by assuming a systematic error of 10\%.
At first, we evaluate the continuum spectrum using Negative
and Positive Exponent~(NPEX) model, as
\[
F_{\rm NPEX} (E;A_{\rm n}, A_{\rm p}, \alpha, \beta, T) =
(A_{\rm n}E^{-\alpha} + A_{\rm p}E^{+\beta}) \cdot exp(-\frac{E}{T}).
\]
where A$_{\rm n}$, A$_{\rm p}$ are normalization and  $\alpha$, $\beta$ 
are photon index,
T is a cut-off energy and E the photon energy.
This model successfully reproduces the subtle curvature change in the binary X-ray pulsar continuum~(Makishima et~al. 1999, Mihara 1995). 
We can evaluate the parameters from the fitting and 
the reduced ${\chi}^2$ is 2.69~(179 degrees of freedom(d.o.f.)) for JEM-X and IBIS
simultaneous fitting.
The obtained reduced ${\chi}^2$ suggests that 
the single NPEX continuum cannot explain the JEM-X and IBIS spectrum,
and we can clearly find an excess of the residual at about $\sim$ 35~keV.
Then we add the CRSF model, as
\[
F(E) = AE^{-\Gamma} exp\{ - \frac{D(WE/E_{\rm c})^2}{(E-E_{\rm c})^2 + W^2}\}
\]
where F(E) is photon number at energy E, A is normalization, $\Gamma$ is 
photon index, while E$_{\rm c}$, D, and W are the 
energy, depth, and width of the cyclotron resonance line.
The exponential factor represents the classical electron cyclotron
resonance scattering.
After adding the CRSF model into NPEX continuum, 
the fit is improved and the reduced ${\chi}^2$ is 1.60~(176 d.o.f.).
The energy E$_{\rm c}$ is 35.6$\pm$ 0.3~keV and depth D is 1.1$^{+0.8}_{-0.1}$. 
 
\begin{figure}
\centering
\includegraphics[width=0.70\linewidth,angle=270]{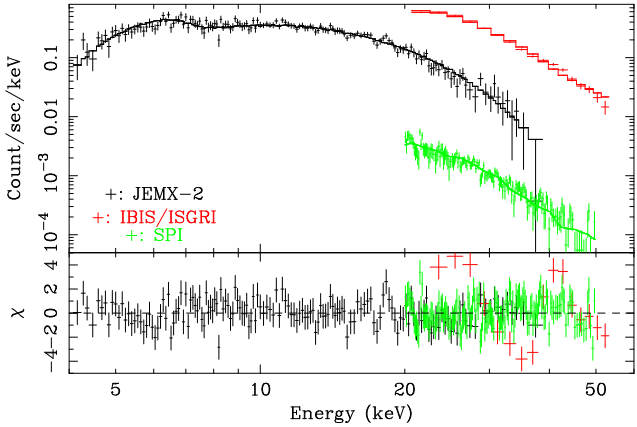}
\caption{The ISGRI, JEM-X and SPI simultaneous fitting of GX301-2 
with a detector response matrix.}
\label{fig:crab_fit}
\end{figure}

\subsection{Luminosity dependence of the cyclotron line}
\begin{figure}
\centering
\includegraphics[width=0.73\linewidth,angle=90]{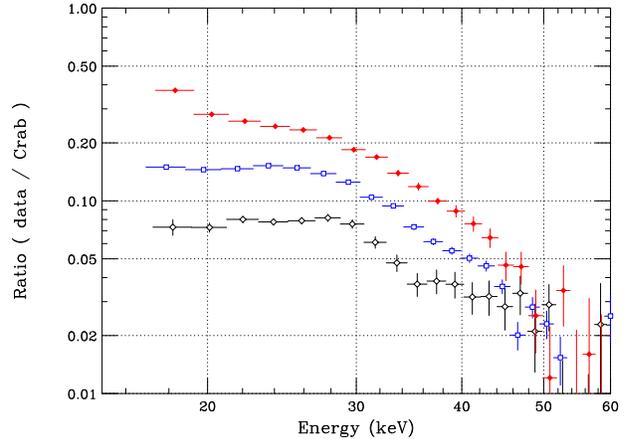}
\caption{Luminosity dependence of the X-ray spectrum, upper ~(red)
, middle ~(blue) and bottom plot~(black) shows the spectrum 
which count rate in the range of 15--40~keV  is defined as $>$ 20~ph/s/bin, 20 $>$ {\it cps} $>$ 10, $<$ 10~ph/s/bin.} 
\label{fig:crab_rate}
\end{figure}

Finally, in order to study the luminosity dependence of the 
CRSF in GX301, we extract the spectra from three different 
surface brightness separately, which is defined as {\it cps} 
$>$20 count/s, 20$>$ {\it cps} $>$ 10 counts/s, cps $<$ 
10 counts/s, in the energy range of 15--40~keV respectively.
Figure 5 shows the spectrum of three different surface
brightness.
We can clearly see that the depth of the CRSF increases as the 
luminosity decreases.
Because of the low statistics and unreliable response matrix,
we do not show more detailed result in figure \ref{fig:crab_rate}.

\section{Twin pulsars}
1E1145.1-6141 and 4U1145-691 locates very close each other, $\sim$ 
20$^{\prime}$,
and are known as "{\it twin pulsar}".
With {\it INTEGRAL}, we can spatially separate the sources with JEM-X and possibly IBIS.
From 1E1145.1-6141, we marginally find the CRSF  at about 20~keV by the Crab-ratio analysis, but the detection is still marginal.
From H1145-619, CRSF is not detected in the range of 5--80~keV.

\begin{figure}
\centering
\includegraphics[width=0.73\linewidth,angle=90]{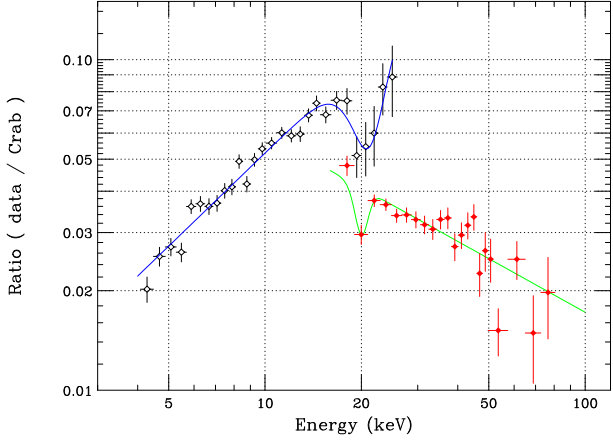}
\caption{The ISGRI~(red/green) and JEM-X~(black/blue) Crab ratio fitting of 1E1145.1--6141.}
\label{fig:1e1145}
\end{figure}

\begin{figure}
\centering
\includegraphics[width=1.0\linewidth]{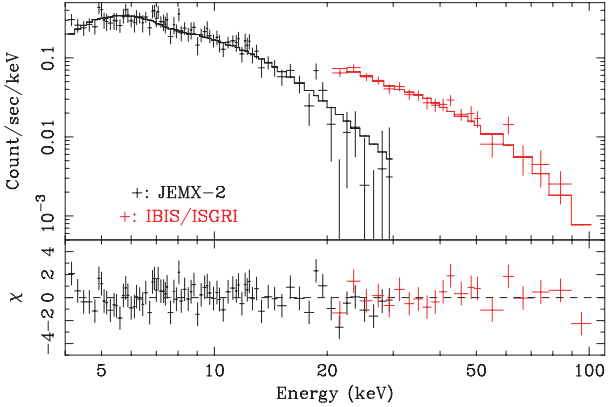}
\caption{The ISGRI and JEM-X spectrum fitting of 4U1145-619.}
\label{fig:4u1145}
\end{figure}

\section{Cen X-3}
\begin{figure}
\centering
\includegraphics[width=0.73\linewidth,angle=270]{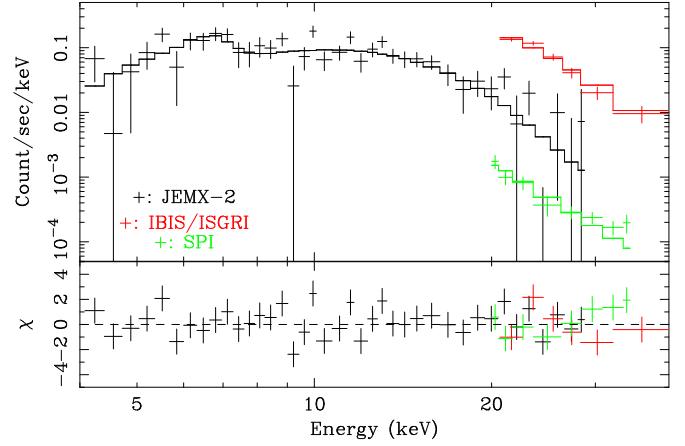}
\caption{Cen X-3 spectrum fitting with ISGRI, JEM-X and SPI.}
\label{fig:cenx-3}
\end{figure}

One of our primary purpose of Cen X-3 observation is to search for the 
higher harmonic. However the Cen X-3 was very dim with a flux of $\sim$ 15~mCrab 
in 15-100~keV energy band.
Figure \ref{fig:cenx-3} shows the JEM-X,~IBIS, and SPI simultaneous 
fitting of the Cen X-3 observation.
The data are successfully reproduced by the NPEX model,
but we fail to detect the CRSF at about 20~keV,
which is reported by Santangelo et al. 1998.
Further deep observation is important.
\section{Summary}
We observed the Cen-crux region in order to search for the cyclotron resonance
scattering features with {\it INTEGRAL}.
During 200~ksec observation, we have detected 4 X-ray binary pulsars.
For GX301-2, the CRSF is clearly detected at about $\sim$ 37~keV 
in the phase averaged spectrum and 
we found that the depth of the CRSF depends on the X-ray luminosity.
The center energy of the CRSF is consistent with the {\it Ginga} 
result~(Mihara 1995).
The {\it twin pulsar} is spatially separated by JEM-X and IBIS/ISGRI.
For 1E1145.1-6141, one of the pulsar twins, we marginally detected
the CRSF at about $\sim$ 20~keV. This is a first detection if the data 
are statistically significant.
Cen X-3 was very dim and resultantly the CRSF is not detected significantly.
Further detailed analysis is being expected after a new release of OSA.

\end{document}